# Sociotechnical Management Model for Governance of an Ecosystem

Antonio J. Balloni[1], Adalberto Mantovani Martiniano de Azevedo[2] and Marco Antonio Silveira[3]

[1]Researcher at Center for Information Technology Renato Archer – CTI/Br
antonio.balloni@cti.gov.br
[2]Federal University of ABC
adalberto.azevedo@ufabc.edu.br
[3]Researcher at Center for Information Technology Renato Archer – CTI/Br
Marco.silveira@cti.gov.br

## ABSTRACT:

*This is an opinion paper regarding a proposal of a model for a Ecosystemm Governance. In the globalized world the importance of Information Systems (IS) and Information Technology (IT) become increasingly relevant regarding the requirements imposed by competition. Both the knowledge of the business as the rapid flow of information are fundamental for a enterprise decision making. Whereas the basic definition of IT = hardware + software, i.e. , tools that has been used to create, store and disseminate data and information in the creation of knowledge, and IS = IT + People + procedures that collect, process and disseminate the information to support decision making, coordination, control, analysis and visualization in the organization [01], it makes implicit the understanding of IS is essential to create competitive companies, to manage global corporations and provide customers with products and services of value. In this work we are correlating IS with the governance of management of an ecosystem.*

*Yet, as IT is redefining the foundations of business, then the customer service, operations, strategies of product marketing and its distribution and even the knowledge management (KM) depends very much, or sometimes even completely, on the IS. The IT and its costs have become a part of day-to-day business [02]. In order to meet this complexity of business needs, today is not possible to disregard the IT and its available resources, which makes very dificult to draw up IS. Therefore, the perspective view of the Sociotehcnical Aspects of an IS are directly concerned with governance and the model proposed regarding an ecosystem.*

*Finally, whereas the summary above, the main objective of this opinion paper is to propose the guidelines for a Sociotechnical Management Model of Governance for an Ecosystem.*

**Keywords:** *Governance, Ecosystem Management, Sociotechnical System, Information Systems, Information Technology.*

# I. INTRODUCTION

**Why Sociotechnical Management Model for Governance of an Ecosystem?** A simple answer would be: because in this globalized world modern organizations need to understand that governance is, and always has been, the tonic of the management and, within this context due to the increasing importance of organizational ecosystems, the creation of a model for a better understanding of all relationships in this system is a must. Yet, the comprehension of an ecosystem must be considered as an important new functional area for any companies' operations. That said, an understanding of the sociotechnical aspects of the governance for an ecosystem is essential for the success or failure of organizations and enterprises and, consequently, the success or failure of a specific ecosystem, be it isolated or open ecosystem (figures 3 and 5).

Nowadays, only with the governance of an ecosystem is possible to create healthy and competitive enterprises, to manage global corporations, to provide customers with products and services of value and, mainly to manage the knowledge as a factor of production. In an sociotechnical approach governance is the tonic of the management

**The Sociotechnical Model for Governance of an Ecosystem** is based on the premise that knowledge of the sociotechnical aspects of an ecosystem is essential to create competitive businesses or systems, to manage global corporations, to provide customers with products and services of value, to manage knowledge as a factor of production and, above all, by the fact that governance is the tonic of management as presented in this work.

# II. ECOSYSTEM OF AN INFORMATION SYSTEM

The ecosystemof an IS comprises the definition of IS presented in the summary, with their input (data), processing (form data to information) and output (the spread out the information for management use in the generation of new knowledge or decisions making). The integration from IS + Organization (business strategy, rules and process) + Environment (suppliers, competitors etc.), represents an expanded vision of an ecosystem, that aims to recognize the actors and their relationships, that are the formative elements of an information ecosystem. This expanded vision is necessary for building the proposed model: the sociotechnical governance of an ecosystem.

# III. THE SOCIOTECHNICAL SYSTEM

A sociotechnical system aims to study the interaction of the human element (organizations) with the technologies surrounding them, seeking to understand how people search, obtain, evaluate, share, classify and make use of the information and news forms of interaction provided by the IT. With a better understanding of these factors is expected an increase of the competitive advantage by the organizations.

Aiming to a better understanding among the sociotechnical system, management of information technology and the communication via corporate governance in an ecosystem, the figure 01 presents the interaction between such organizations (human element) with IT (IS): a sociotechnical system is comprised by an social system (organizations) and Technical System (IS)

# The Interdependence between Organizations and Information Systems

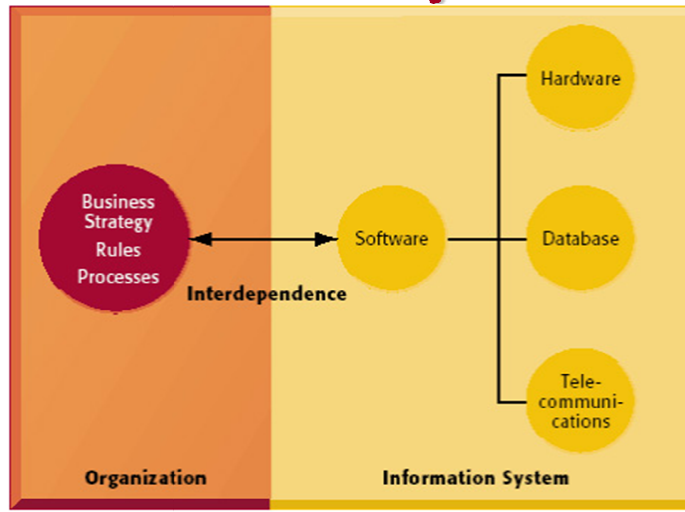

Figure 01: The interdependence between a social system (organization) and a technical system (IS). There is interdependence between organizations and IS. [03, 04].

- <u>The social system.</u> Organization, figure 1 - comprises the employees (at all levels) and the knowledge, skills, attitudes, values and needs they bring to the work environment as well as the reward system and authority structures that exist in the organization. [05].

- <u>The technical system.</u> Information System, figure 1- comprises the devices, tools and techniques needed to transform inputs into outputs in a way which enhances the economic performance of the organization. [05].

- <u>Interdependency.</u> The basis of the sociotechnical approach is: the fit is achieved by a design process aiming mutual optimization of all systems (socio + technical), figure 2. Any organizational systems will maximize performance only if the interdependency of these systems is explicitly recognized. This approach prevents a purely technological IS. This interdependency is the basic premise for the creation of the concept "Isolated model of an ecosystem of Business System", as explained in section IV (figure 03).

# A Sociotechnical Perspective on Information Systems

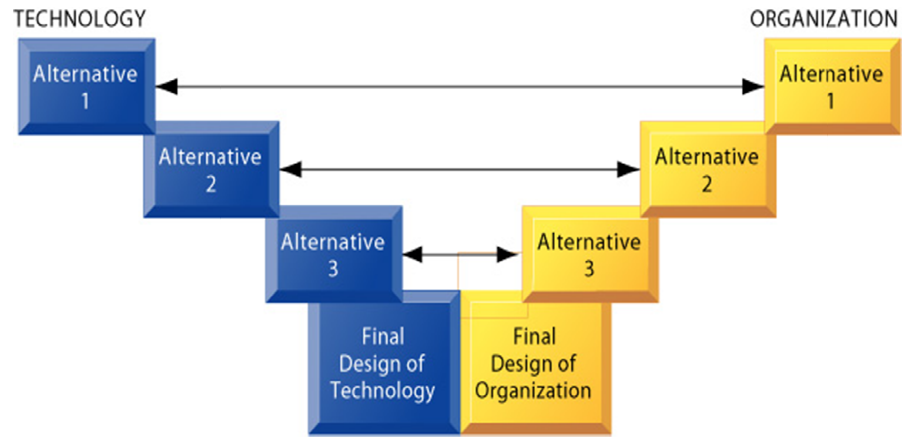

Figure 02: Sociotechnical perspective on Information System (IS): the performance of an IS is optimized when both the technology and the organization mutually adjust to one another until a satisfactory fit is obtained. [05,06]. This interdependency is the basic premise for the creation of the concept "Isolated model of an ecosystem of Business System", as explained in section IV (figure 04).

However, to take all these factors in account is not an easy and may be not a feasible task. Sociotechnical systems, understood as complex ecosystems, are Self-Organizing Holarchic Open Systems (SOHO) [07]. According to the authors, these complex systems are arrangements of mutable interactions between their components, related to flexible hierarchies, leading to the constant reconfiguration of one organizational state to the other.

In order to deal with the uncertainties inherent to the Information Systems, adaptative learning capabilities are required, as well as in political elements of the ecosystem, such as its hierarchical and governance foundations.

By managing its organizational ECO, an organization is able to acquire external competences and resources of other organiza-tions (be it data, information, concepts, knowledge or wis-dom).This is especially important in cases where organizations aim to reach costly and complex objectives, that require the conjointy work of several organizations with different competences and resources, such as research and development arrangements. A participatory organizational ECO approach can be used as a guide for balanced organizations that aim to benefit from its ECO´s resources, taking as guidance the following principles that should be adopted for improving the management of organizational ecosystems: 1. Open and lateral dissemination of know-how; 2. Freedom to voice contrary opinions; 3. Frequent face-to-face interaction; 4. Transformation of tacit knowledge into explicit knowledge; 5. Formal and informal organizational support mechanisms to manage the organizational ECO.

## IV. ECOSYSTEM VS. ORGANIZATIONAL SUSTAINABILITY

An ecosystem is a network of plants, animals and people, micro-organisms and their nonliving environment. The well-being of the people is intimately linked to the health of ecosystems, because these systems depend on them for various resources and services, such as wood, food and water, and to regulate the climate. It is important to protect all components of an ecosystem due to the interdependence of all its elements [08]. In a similar way to a natural ecosystem, an organizational ecosystem is a network of organizations and its environment which are in continuous transformation (a very know rule of nature, the Lavoisier's Law, states that: "In Nature nothing is gained, nothing is lost, everything is transformed "). Thus, it can be said that the well-being organizational is closely related to the health of the ecosystems where is immersed that organization. The health of the ecosystem depends on its components and strategies [09]. There are ecosystems where predatory ("wolf") strategies are the rule (what is no good for ecosystems health on the long term). On the other sides, ecosystems where components adopt more cooperative ("beaver") strategies seem to be more healthy in the long term, since its members are concerned with the system stability, rather than to grab short term benefits.

**The management of an ecosystem happens via governance. Therefore, governance is the tonic of management: it does not stifle or binds an ecosystem**. Governance structures are created by ecosystems members- depending on the conditions of the environment and on the relationships established inside the ecosystem. For instance, an organizational ecosystem can be constituted by several firms (suppliers, retailers, etc.) that work together to produce and sell a final good, under the command of one company (the "keystone", as defined by [09]). This "keystone" company, by means of its strategy, guides the efforts of all its partners and keeps the ecosystem healthy, by providing directions on production and marketing efforts by mastering key elements such as the product´s design and the business plan elaborated to take the maximum profit of it.

These structures must be designed to dynamically adapt to the changes of the organizational ecosystems and its environment (sociotechnical systems, Section III), Figures 03, 04 and 05. This will be clarified in the next sub-section, IV.1 - regarding an isolated ecosystem model from a business environment.

## IV. 1 - ISOLATED ECOSYSTEM MODEL OF A BUSINESS ENVIRONMENT

Regarding the Lavoisier's Law (Section IV), the figure 03 presents the model "An Isolated Ecosystem Model of a Business Environment". This model comprises the concepts of IS and sociotechnical system and, it is an abstraction of the concept regarding the interactions between or among ecosystems. Example: consider an isolated ecosystem with a given Research, Development and Innovation or product development. Since the ecosystem is isolated (no interaction between its borders), then this ecosystem will die, and consequently, the whole chain of interaction of a broad ecosystem within which the isolated ecosystem is immersed will also die!

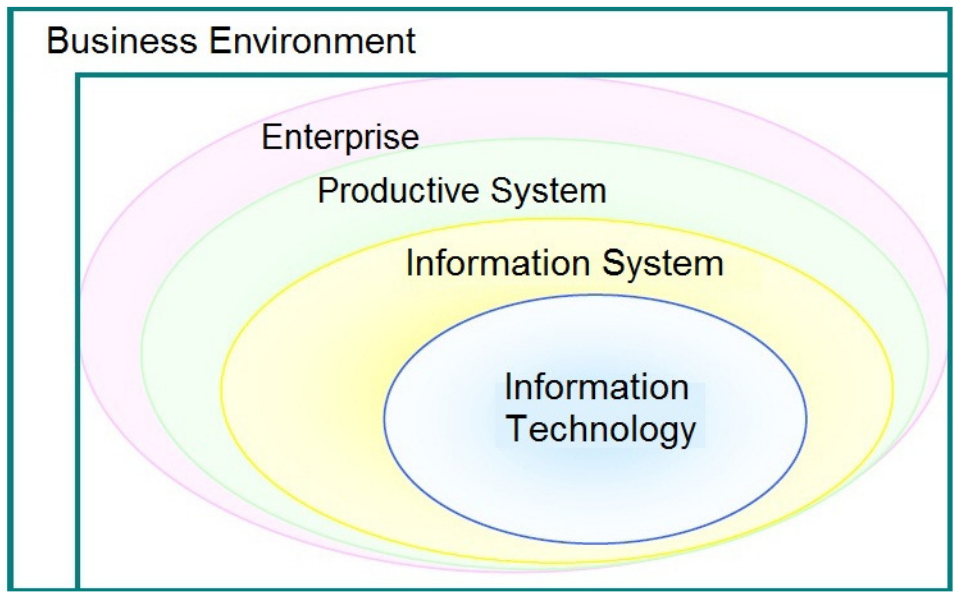

Figure 03: Isolated Ecosystem Model of a Business Environment. Adapted from [10].

Therefore, an isolated ecosystem, figure 03, will only be sustainable if it fulfills the condition of Organizational Sustainability: state that provides in medium to long term, the best systemic gains in function of the best conditions for rational use of natural resources, minimization of environmental impacts and human development. In others words, conditions to maintain, in the long term, the health of the ecosystem, in a sustainable way. Section IV.2 presents the Model for Organizational Sustainability from a Business Environment.

## IV.2 - MODEL FOR ORGANIZATIONAL SUSTAINABILITY FROM A BUSINESS ENVIRONMENT

Figure 04 presents a model of organizational sustainability [11]. The isolated ecosystem, shown in figure 03, will be sustainable only if complies with the criterion of organizational sustainability tripod, shown in figure 04.

The intersection between economic performance, environmental balance and society needs, (figure 04) shows that governance ( ⭐ ) is the tonic of management. Any system will be able to maximize its performance if and only if the interdependence (between those systems, figure 01) has been explicitly recognized, and governance structures designed to manage these interdependent relationships.

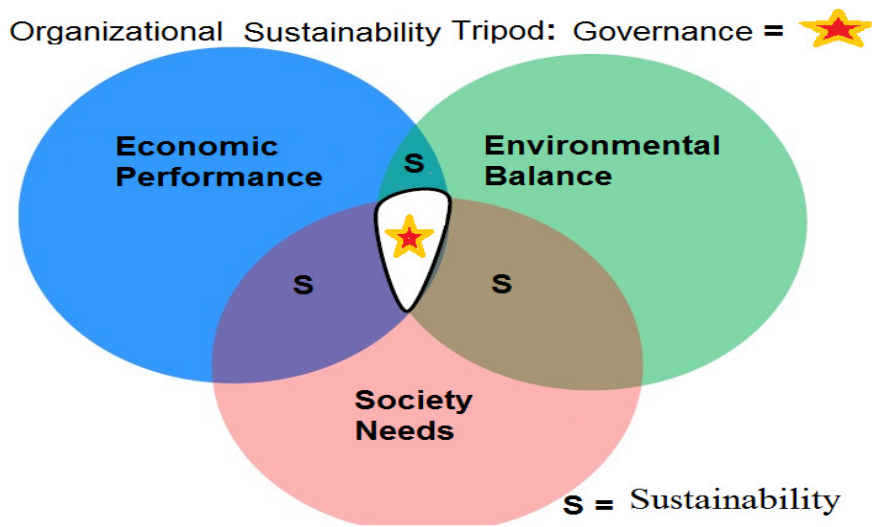

Figure 04: Model of Organizational Sustainability Managed by Governance. Governance is the Tonic of Management in the Tripod of Sustainability [11].

According to [12] , the term sustainability comprises dynamic cultural, economic, and biophysical systems and its association with landscape elements, such as quality of life for humans individuals or collectives. In a sustainable way, human activities "… do not threaten the integrity of the self-organizing systems that provide the context for these activities".

To further clarify this definition of sustainability, we need a complementary definition for integrity. The integrity of an ecosystem that integrates biophysical, cultural, and economic systems depends on the ecosystem's retain of ots own complexity, capacity for self-organization and diversity, [12].

The health of an ecosystem depends, from our point of view, on the balance between the three dimensions of sustainability. That is to say, a healthy ecosystem governance structures must be designed in a way that minimizes aggressions to the natural environment and maximizes societal benefits, without ever loosing sight of its economic performance. In order to do this, ecosystems management must take into account the variables associated with different levels of the organization: the ecosystem itself (that is, the collection of public and private stakeholders related to the ecosystem), the organization immersed on the ecosystem and, last but not least, the individuals (human capital) that are at the core of any human organization.

## IV.3 – "Ecosystems are complex dynamic systems. [13]

Ecosystems are composed of many mutually interdependent parts operating in dynamic, co-evolutionary trajectories. They are not static, and do not necessarily tend toward equilibrium. Parts interact with other parts in rich, multiple (and often poorly understood) ways, so that the arrows of causation–what action causes what effect–often point in many directions simultaneously, some in self-reinforcing chains, some held in check by others tending in the opposite direction (positive and negative feedback loops).

Ecosystems exhibit nonlinearity in many dimensions. That is to say, the effects of many actions are discontinuous. For example, there are numerous natural threshold effects, as well as complications caused by co-causation and synergistic interactions among multiple factors operating along multiple complex chains of causation, often incorporating both positive and negative feedback loops simultaneously. As a result, small inputs can sometimes result in large, and often partially or even wholly unpredictable consequences for other parts of the system, and for the system as a whole. As with other complex systems (as an organization), even if our understanding of the individual components and their operational principles is relatively complete, our understanding of the trajectory of the entire system qua system, and the ultimate effect that certain inputs will have on the system as a whole and its individual component parts, may be quite limited. Therefore, in any organization within its Isolated Ecosystem, figure 03 (Isolated Ecosystem Model of a Business Environment), although we can do our best to plumb the deep organizational principles and to monitor conditions carefully to give us as complete a picture as possible, we need to expect and prepare for surprises (Input, processing and output). This is the sense in which an ecosystem is a complex, nonlinear dynamic system…" [13]

## V - SOCIOTECHNICAL MODEL FOR GOVERNANCE OF AN ECOSYSTEM

Figure 5 is an adaptation of Vendrametto [14] but, now, as a model of Sustainable Ecosystem towards an Open Ecosystem (not isolated). This integrated set corresponds to the **Sociotechnical Management Model for Governance of an Ecosystem.**

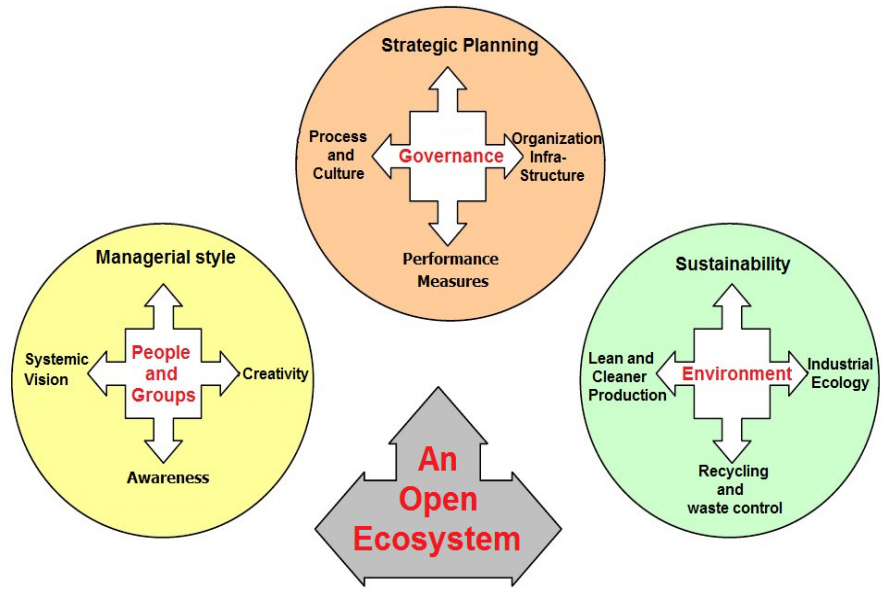

Figure 5: Model of Sustainable Ecosystem towards an Open Ecosystem (not isolated). An adaptation from [14]: "Implementação Metodológica do Modelo de Governança do Conhecimento para a Sustentabilidade no Senai - Serviço Nacional de Aprendizagem Industrial"

# VI – PERSPECTIVE FOR MANAGEMENT OF AN ECOSYSTEM

With the correct combination among the management of IS and IT, the business processes and administrative abilities (persons, groups and environment), the return on investment (ROI) regarding the implementation of the Sociotechnical Management Model for Governance of an Ecosystem may be effective. Governance is the Tonic of Management of an Ecosystem!

The governance of an ecosystem can help companies to improve their business process and decision making efficiency and effectiveness, harmoniously integrated in this ecosystem and, within a globalized market in broader transformation [1,2 ] (not isolated).

The "Model of Sociotechnical Management for Governance of an Ecosystem" presented in this opinion paper, integrates completely with the model of Information Architecture developed by Balloni [1,2 ] and here adapted as the Model of the Architecture of an Ecosystem, figure 6.

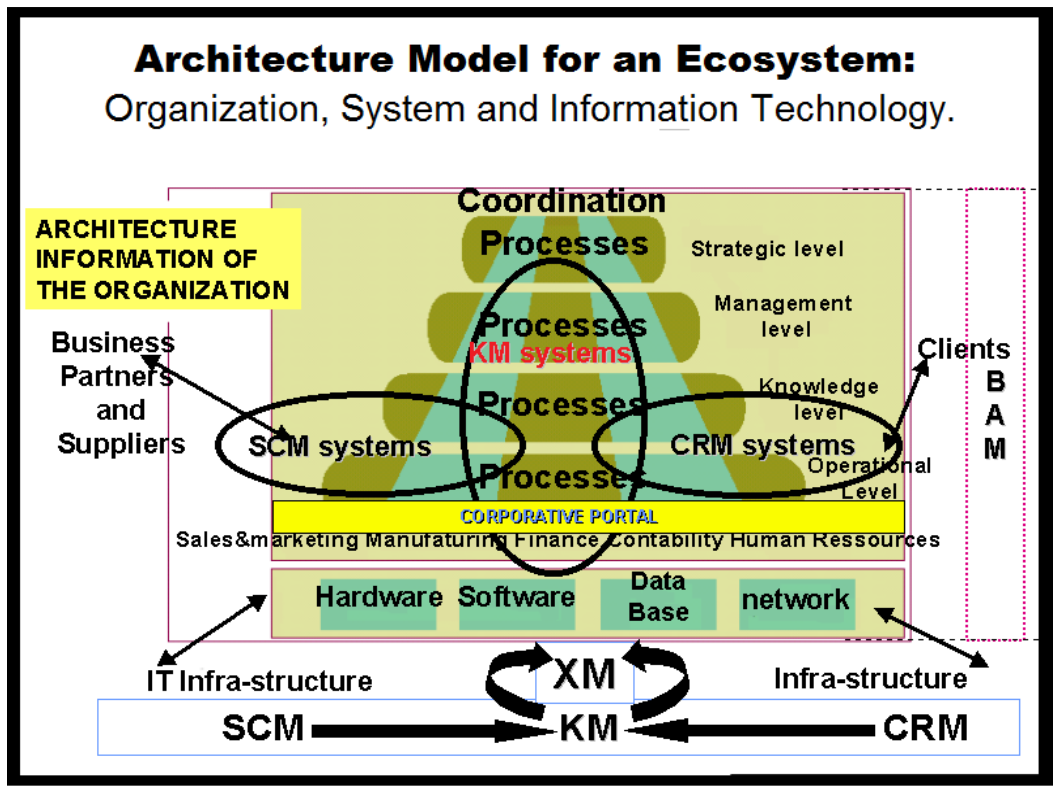

Figure 6: Model of Architecture of an Ecosystem. The Architecture of an Ecosystem deals with the particular project of an ecosystem towards a chain or specific niche of market, [01,02]. The IT infrastructure is the platform on which organizations may build its specific IS (hardware, software and connections between systems).

Managers must know to arrange and coordinate the various applications of IS (SCM, SCRM, SKM), to meet the enterprise needs at the: level of organizational (strategical, managerial, knowledge and operational), the organization as a whole (organization architecture) and the architecture of the ecosystem (XM and BAM). Managers also should compose in a unique way the interconnection of an organization with other organizations (interorganizational systems).

SSCM: information system for Supply Chain Management;
SCRM: information systems for Customer Relationship Management,
SKM: information system for Knowledge Management and
BAM: business Activity Monitoring.
XM: Integrated Information System

To maximize the benefits of the governance of an ecosystem is necessary to plan the Architecture of this Ecosystem and, this is the great managerial challenge, i.e., to create a uniform sociotechnical system uniform in which everyone is using similar processes and information: integration of key business processes of this ecosystem and improvement of the coordination, efficiency and decision making.

The main function of the BAM is to be focused on business processes, providing real-time information regarding the key indicators of the organization (ecosystem) performance.

The Corporate Portal allows the ecosystem the necessary interactivity and accessibility to those key information (certified by digitals publics keys).